\documentclass{aa}

\authorrunning{Caunt \& Tagger}
\titlerunning{Numerical simulations of the AEI}

\usepackage{graphics}
\usepackage{epsf,epsfig}
\usepackage{amssymb}
\def\ltsima{$\; \buildrel < \over \sim \;$}
\def\simlt{\lower.5ex\hbox{\ltsima}}
\def\gtsima{$\;\buildrel>\over\sim\;$}
\def\simgt{\lower.5ex\hbox{\gtsima}}

\begin{document}
%
%
\newcommand{\cm}{\,{\rm cm}}
\newcommand{\mm}{\,{\rm mm}}
\newcommand{\km}{\,{\rm km}}
\newcommand{\cmcube}{\,{\rm cm^{-3}}}
\newcommand{\gcmcube}{\,{\rm g\,cm^{-3}}}
\newcommand{\kgcmsqr}{\,{\rm kg\,cm^{-2}}}
\newcommand{\dyn}{\,{\rm dyn}}
\newcommand{\erg}{\,{\rm erg}}
\newcommand{\Jy}{\,{\rm Jy}}
\newcommand{\Jyb}{\,{\rm Jy/beam}}
\newcommand{\kms}{\,{\rm km\,s^{-1}}}
\newcommand{\mJy}{\,{\rm mJy}}
\newcommand{\mJyb}{\,{\rm mJy/beam}}
\newcommand{\K}{\,{\rm K}}
\newcommand{\kpc}{\,{\rm kpc}}
\newcommand{\Mpc}{\,{\rm Mpc}}
\newcommand{\mG}{\,{\rm mG}}
\newcommand{\mkG}{\,\mu{\rm G}}
\newcommand{\MHz}{\, {\rm MHz}}
\newcommand{\Msol}{\,{\rm M_\sun}}
\newcommand{\p}{\,{\rm pc}}
\newcommand{\radm}{\,{\rm rad\,m^{-2}}}
\newcommand{\s}{\,{\rm s}}
\newcommand{\yr}{\,{\rm yr}}

\thesaurus{02.13.2; 02.09.1; 02.01.2; 11.10.1}

\title{Numerical simulations of the Accretion-Ejection Instability in
magnetised accretion disks}

\author{S.~E. Caunt and M.\ Tagger}
\institute{DSM/DAPNIA/Service d'Astrophysique (CNRS URA 2052), C.E.A.
Saclay, 91191 Gif sur Yvette, France}

\maketitle

\begin{abstract}

The Accretion-Ejection Instability (AEI) described by Tagger \& Pellat
(\cite{TaggerP99}, hereafter TP99) is explored numerically using a
global 2d model of the inner region of a magnetised accretion disk.  The
disk is initially currentless but threaded by a vertical magnetic field
created by external currents, and frozen in the flow.  In agreement with
the theory a spiral instability, similar in many ways to those observed
in self-gravitating disks, develops when the magnetic field is, within a
factor of a few, at equipartition with the disk thermal pressure.
Perturbations in the flow build up currents and create a perturbed
magnetic field within the disk.  The present non-linear simulations give
good evidence that such an instability can occur in the inner region of
accretion disks, and generate accretion of gas and vertical magnetic
flux toward the central object, if the equilibrium radial profiles of
density and magnetic flux exceed a critical threshold.


\end{abstract}

\section{Introduction}

Magnetic fields in accretion disks have been seen over the last decade
to strongly effect the structural and dynamical properties of
accretion disks as compared to the well defined standard steady-state,
thin disk models (Shakura \& Sunyaev \cite{ShakuraS73}, Pringle
\cite{Pringle81}, Frank, King \& Raine \cite{FrankKR92}). Since the
early 1990s, numerical simulations of accretion disks have
increasingly shown huge deviations from the smoothly varying,
symmetrical (azimuthally and vertically), time-independent accretion
disk models.

Although the magnetic field has been assumed to be of importance in
generating instabilities within the disk (leading to turbulent
transport), this has essentially been modeled as turbulent viscosity in
the classical prescription $\nu=\alpha c_s h$ where $\nu$ is the
viscosity, $c_s$ the sound speed, $h$ the disk height and $\alpha$ a
dimensionless parameter.  This then allows for the outward radial
advection of angular momentum and inward radial transport of matter.
However, this alone is insufficient to explain the time-variability of
accretion disks such as relatively long timescale (hours to days) dwarf
novae outburst of cataclysmic variables ({\em e.g.} Cannizzo \& Mattei
\cite{CannizzoM92}) or more rapid (of the order of seconds) QPOs ({\em
e.g.} Swank {\em et al} \cite{SwankCMT97}, Muno{\em et al.} \cite
{Muno99}, Sobczak {\em et al.} \cite{Sobczaketal00}).  The time
variability of these systems has always been shown to be associated with
the dynamics of the accretion disk itself - a feature that is not
included within standard hydrodynamical treatments.

Since the discovery of the importance of the magneto-rotational (MRI)
or Balbus-Hawley (BH) instability (Velikhov \cite{Velikhov59},
Chandrasekhar \cite{Chandra60},\cite{Chandra61}) in the context of
accretion disks by Balbus \& Hawley (\cite{BalbusH91}), and its
numerical demonstration by Hawley \& Balbus (\cite{HawleyB91}), a
number of local 3d simulations have been performed ({\em e.g.}
Brandenburg {\em et al.} \cite{BNST95}, Hawley {\em et al.}
\cite{HawleyGB95}, Stone {\em et al.}  \cite{StoneHGB96}, Hawley {\em
et al.} \cite{HawleyGB96}, Hawley \cite{Hawley00b}). These all
show that turbulent transport is indeed generated through a local
instability.  The resulting transport, if it is measured in terms of
the $\alpha$ prescription, is in the range expected from observations;
however the resulting $\alpha$ parameter is strongly varying, meaning
that the turbulence does not obey the simple prescriptions of the
$\alpha$ model. Other features, such as time variability (Hawley \&
Krolik \cite{HawleyK00}), a vertically varying $\alpha$ (Brandenburg
{\em et al.} \cite{BNST96b}) and vertical structural asymmetries
(Caunt \cite{Caunt98}, Miller \& Stone \cite{MillerS00}), imply that
the magnetic field strongly affects the overall dynamics of an
accretion disk (and in particular its vertical structure and its
connection to the corona).

Jets, associated with accretion disks in Active Galactic Nuclei, X-ray
binaries and Young Stellar Objects are certainly magnetised (Blandford
\& Payne \cite{BlandfordP82}) and the magnetic field at the disk surface
is required to drive the matter across the slow magnetosonic point,
after which centrifugal forces take over (Lovelace {\em et al.}
\cite{LovelaceBC91}).  The formation of jets is also observed to be
connected to the accretion within the disk. In fact one the main
attractions of MHD jet models is that they show that the jet is a very
efficient way to extract angular momentum from the disk, in sharp
contrast with $\alpha$-disk models where the angular momentum is
evacuated outward radially by viscous stresses.  A number of
instabilities other than the MRI have been proposed for accretion disks
({\em e.g.} Parker \cite{Parker79}, Papaloizou \& Pringle \cite{PapP84},
Spruit {\em et al.} \cite{SpruitSP95}); however the recent discovery of
the accretion-ejection instability (AEI) (Tagger {\em et al.}
\cite{TaggerHSP90}, TP99) provides a possible link between radial and
vertical angular momentum transport.

As shown by TP99, the instability occurs close to the inner edge of an
accretion disk for which the quantity $\Omega\Sigma/B_0^2$ increases
radially where $\Omega$ is the angular velocity, $\Sigma$ the surface
density and $B_0$ the magnetic field for which the plasma beta (the
ratio of gas to magnetic pressures) is of a moderate value (around
unity).  The instability appears as a low azimuthal wavenumber spiral
wave, made unstable through the interaction with a Rossby vortex
(associated with the gradient of vorticity) which it generates at
its corotation radius.  This also causes the emission of Alfv\'en waves
from corotation vertically along magnetic field lines, as their
footpoints are twisted.  Hence, in the context of jet formation, the AEI
naturally contains the highly desirable property of linking vertical
propagation directly to the disk dynamics.  Also, unlike the local MRI
which is quenched for moderate fields, the AEI exists in situations for
which the field pressure is comparable to the gas pressure, as may
occur in the inner region of an accretion disk as magnetic field is
advected inwards with matter and accretes towards the central object.

With the view of the low-frequency QPOs observed in X-ray binaries,
in particular those hosting a black hole, the instability is also a
good theoretical candidate (Varni\`ere {\em et al.},
\cite{VarniereRTD00}; Rodriguez {\em et al.}, \cite{RodriguezVTD00}).
One can thus imagine a scenario in which, under initial low field
strength situations the MRI initially acts, advecting the field radially
inward by turbulent transport. If at least part of the vertical
magnetic flux is advected with the gas the field in the inner disk
region builds up, until a certain point is reached where the higher
magnetic pressure quenches the MRI and the AEI sets in.  This `magnetic
flood scenario' (Tagger \cite{Tagger99}) might explain the $\sim$ 30
minutes cycles of the micro-quasar GRS 1915+105 (Mirabel {\em et al.}
\cite{Mirabel98}).

The inner region of accretion disks is increasingly becoming the focus
of 3d numerical experiments.  As computational power increases, global
models (still limited by resolution and radial extent) are now being
produced as well as the well-established local shearing-box models.
Armitage (\cite{Armitage98}) describes a global model that omits
vertical stratification and is limited in both vertical and radial
extent, but illustrates effectively the global development of the
MRI. More recently Hawley (\cite{Hawley00}) and Machida {\em et al.}
(\cite{MachidaHM00}) describe simulations of accretion disks starting
from toroidal configurations ({\em i.e.} a departure from Keplerian
thin disks); at the time of preparation Hawley \& Krolik
(\cite{HawleyK00}) and Hawley (\cite {Hawley00b}) have published
numerical simulations of full accretion disk models.

We note here, and will discuss in our conclusions, that in these 3d
simulations spiral waves are seen to occur close to the inner boundary
of the disk as a result of magnetic stresses; this is very similar to
the features expected from the AEI, and indeed very similar to the
results presented here.

In a very different approach, Stehle \& Spruit \cite{StehleS} have
presented a 2D ($r, \ \phi$) model of a disk.  The disk is considered
as infinitely thin and embedded in vacuum.  Their model is used to
demonstrate the existence of the interchange instability, predicted to
occur in strongly magnetised disks (Spruit \& Taam \cite{SpruitT90},
Spruit {\em et al.} \cite{SpruitSP95}), and associated with the radial
gradient of the magnetic field. Unexpectedly, they find that the disk
is also unstable for lower field strength than predicted.

Our model is very similar to theirs, with numerical characteristics
optimised according to our previous knowledge of the AEI. We also
consider an infinitely thin disk threaded by a moderate vertical
magnetic field: this is justified by the properties of the AEI, which is
essentially constant vertically across the disk.  The disk is also
embedded in vacuum but, in order to separate different physical effects,
we consider in the present paper only configurations where initially the
equilibrium magnetic field is due to external currents.  As discussed
below we use a logarithmic radial grid: this allows us to get a better
resolution in the inner region of the grid, where it is necessary, and
on the other hand to model a much larger radial extent of the disk and
to get rid of unwanted effects of the boundary condition at the outer
radius of the simulation.  We believe that such effects may explain the
stronger instability observed in their work, compared with ours.  This
numerical setup (with, as discussed below, a magnetic potential given by
a Poisson equation) allows us to use well-established methods developed
in the description of self-gravitating disks.

In spite of these differences, our results are quite similar to the
ones they obtain at moderate field amplitude and we believe that,
although the initial conditions differ and the numerical methods are
independent, similar physics are at the heart of the two simulations.

This model allows for the development of the spiral wave and the vortex
-- at the heart of the instability.  The paper is organised as follows:
Section 2 describes the basic model including the fluid equations and
implementation of the magnetic field, Section 3 provides details of the
numerical methods used, in Section 4 we present our results and in
Section 5 our conclusions.



\section{The model}

We present, in cylindrical coordinates, a global two-dimensional model
of the inner region of an accretion disk around a central object.  The
model is 2D in the sense that we solve only for the horizontal
($r,\phi$) components of the velocity, and that all perturbed
quantities are assumed independent of $z$ within the disk.  We assume
that the disk is threaded by a poloidal magnetic field which is
symmetric about the midplane, hence purely vertical at $z=0$.  As
described later, assuming that the disk is of a finite vertical height
and embedded in vacuum, the magnetic field outside the disk (and in
particular at its surfaces) can be derived from a magnetic
potential, which in turn, is calculated from the distribution of
magnetic field within the disk (this is analogous to calculating the
gravitational potential in a self-gravitating disk).  Currents within
the disk can be derived from the jump in the horizontal component of
the field across the disk.  Hence magnetic stresses can be included in
an otherwise purely hydrodynamic model.

The variables we solve for are the radial and perturbed azimuthal
velocity, $u_r$ and $u_\phi$, the surface density, $\Sigma$ (which
comes from the vertical integration of the density, $\rho$) and the
vertical component of the magnetic field, $B_z$.  The magnetic
potential is solved separately as $\Phi_m$. The ``perturbed''
component of the azimuthal velocity is defined as its deviation from
keplerian rotation (so that it contains both the true perturbation and
the equilibrium contribution of the pressure gradient).  By
substituting for the full azimuthal velocity $U_\phi=u_K+u_\phi$,
where $u_K$ is the Keplerian velocity, the axisymmetric component of
$U_{\phi}$ can be enforced more accurately and is less prone to
numerical roundoff errors (encountered from large azimuthal velocities
and gravitational forces) which can lead to anomalous radial motion.

The instability can be successfully described using a 2d model because
the vertical scale height of the perturbations is greater than the disk
thickness (TP99), unlike the Magneto-Rotational Instability (MRI) for
which the vertical wavelength is critical, so that 3d local models are
required to study the full nature of the instability ({\em e.g.}
Brandenburg {\em et al.} \cite{BNST95}, Hawley {\em et al.}
\cite{HawleyGB96}, Stone {\em et al.} \cite{StoneHGB96}).  Hence our
simulations will not show the full turbulent transport of angular
momentum expected within accretion disks, but will however allow us to
isolate the effects of the AEI alone in their inner region,
providing a first approach of the physics of the AEI, and valuable
details for future 3d models.

\subsection{Fluid equations}

The model solves for the momentum, induction and continuity equations
integrated vertically across the disk, in the hypothesis that (as
expected for the AEI) the vertical scale height of the perturbations is
larger than the disk thickness. The momentum equations for $r$ and
$\phi$ are given by
\begin{eqnarray}
   \frac{\partial u_r}{\partial t} =
   -u_r\frac{\partial u_r}{\partial r}
   -\frac{U_\phi}{r}\frac{\partial u_r}{\partial \phi}
   +\frac{u_\phi^2}{r}
   +\frac{2u_\phi u_K}{r} \nonumber \\
\hspace{20mm}
   -\frac{1}{\Sigma}\frac{\partial P}{\partial r}
   +\frac{B_z\Delta B_r}{4\pi\Sigma},
\label{e:momr}
\end{eqnarray}
\begin{eqnarray}
   \frac{\partial u_\phi}{\partial t} =
   -u_r\frac{\partial U_\phi}{\partial r}
   -\frac{U_\phi}{r}\frac{\partial U_\phi}{\partial \phi}
   +\frac{u_r U_\phi}{r} \nonumber \\
\hspace{20mm}
   -\frac{1}{\Sigma r}\frac{\partial P}{\partial \phi}
   +\frac{B_z\Delta B_\phi}{4\pi\Sigma},
\label{e:momphi}
\end{eqnarray}

\noindent where $\Sigma$ is the surface density, $U_\phi=u_K+u_\phi$ and
$u_K=\sqrt{GM/r}$ is the Keplerian velocity. Since pressure
stresses are relatively unimportant, we assume for simplicity an
isothermal equation of state so that the pressure gradient only changes
as a result of advection of fluid density.  The vertically integrated
pressure, $P$, is given by
\begin{equation}
P=c_s^2 \Sigma,
\end{equation}

\noindent where $c_s$ is the sound speed.  The terms $\Delta B_r$ and
$\Delta B_\phi$ represent the currents associated with the jumps in
radial and azimuthal magnetic field across the surfaces of the disk.
Assuming symmetry conditions apply then $\Delta B_r = B_r^+ - B_r^- =
2B_r^+$ where the labels + and - note the values at the upper and
lower disk surfaces, and similarly $\Delta B_\phi = 2 B_\phi^+$.

The continuity equation, integrated vertically, gives
\begin{equation}
   \frac{\partial \Sigma}{\partial t} =
   -\frac{1}{r}\frac{\partial}{\partial r}(\Sigma r u_r)
   -\frac{1}{r^2}\frac{\partial}{\partial \phi}(\Sigma r U_\phi).
   \label{e:cont}
\end{equation}

\noindent Similarly the vertical component of the induction
equation takes the form:
\begin{equation}
   \frac{\partial B_z}{\partial t} =
   -\frac{1}{r}\frac{\partial}{\partial r}(B_z r u_r)
   -\frac{1}{r^2}\frac{\partial}{\partial \phi}(B_z r U_\phi).
   \label{e:ind}
\end{equation}

The identical form of the continuity and induction equations results
from flux freezing in the ideal MHD limit, {\em i.e.} from the fact that
both surface density and vertical magnetic flux are conserved in each
tube threading the disk.  It will be used below in the calculation of
the magnetic potential.

\subsection{Magnetic potential}
\label{s:magpot}

Equations \ref{e:cont} and \ref{e:ind} show that the horizontal
advection of the vertical magnetic field has an identical form to the
advection of density in the limits of a vertically averaged model.
Another identical relationship can be derived from assuming that the
disk lies in vacuum.  With this condition the absence of currents
outside the disk allows one to describe the perturbed magnetic field
{\em outside the disk} as the gradient of a magnetic potential,
$\Phi_M$.  Following the derivation in TP99, we describe the field {\em
outside the disk} as:
\begin{equation}
\vec B = -{\rm sign}(z)\vec\nabla \Phi_M,
\end{equation}
\noindent so that both $B_{z}$ and $\Phi_{M}$ are even in $z$, while
$B_{r}$ and $B_{\phi}$ are odd.  In vacuum the divergence-free field
condition gives
\begin{equation}
\nabla^2 \Phi_M=0.
\end{equation}
On the other hand symmetry gives:
\begin{equation}
     \frac{\partial}{\partial z}\Phi_{M}(z=0^+)=-\frac{\partial}{\partial
     z}\Phi_{M}(z=0^-)=-B_{z}^D,
\end{equation}
where $B_{z}^D$ is the  perturbed value of $B_{z}$ in the disk. Thus we
can write throughout space:
\begin{equation}
\nabla^2\Phi_M = -2B_z^D\delta(z),
\label{e:magpois}
\end{equation}
\noindent where $\delta(z)$ is the Dirac function. The form of this is
identical to the standard Poisson equation for the gravitational
potential of a self-gravitating disk:
\begin{equation}
\Delta \Phi = 4\pi G\Sigma\delta(z).
\label{e:gravpois}
\end{equation}

\noindent Hence, from the similarity between the Poisson equations
on one hand, and on the other hand the similarity between the continuity
equation (\ref{e:cont}) and the induction equation (\ref{e:ind}), we can
solve for the magnetic potential using the classical Poisson kernel and
more generally the methods devised in the simulation of self-gravitating
disks (Binney \& Tremaine \cite{BinneyT87}), as described in Section
\ref{s:numpois}.  The integration of the potential from individual field
lines within the disk yields an expression for the total magnetic
potential as:
\begin{equation}
\Phi_M(r,\phi)= \frac{1}{2\pi}\int_{r_1}^{r_2}\int_0^{2\pi}
\frac{B_z^D(r',\theta)r' dr' d\theta}{\sqrt{r^2 + {r'}^2 -2rr'\cos\theta}}.
\label{e:fullpot}
\end{equation}
This similarity explains that, as found by Tagger {\em et al.}
(\cite{TaggerHSP90}), the waves we will describe are very similar to
the spiral waves of self-gravitating disks. An important difference,
however, is the negative sign in Equation \ref{e:magpois}, meaning
that the magnetic field plays the role of a negative self-gravity: its
presence ``stiffens'' the plasma against compression, whereas
self-gravity helps it.

\subsection{Initial Conditions}

For the models presented here we assume, in the context of the
accretion disk of a black hole in an X-ray binary, that the central
object has a mass of $M=10M_\odot$ and the disk extends from between
$r=1,000\km$ to $r=50,000\km$ from the origin at $r=0$ with a constant
aspect ratio of $\epsilon=h/r=0.1$ where $h$ is the disk thickness.
The only critical parameters, however, are $\epsilon$ and the ratio
$\beta$ of thermal to magnetic pressure, so that similar results would
apply in disks of similar aspect ratio but very different dimensions,
such as the accretion disks of young stellar objects or Active Galactic
Nuclei.

The sound speed within the disk is calculated directly from the
thin-disk approximation ({\em e.g.} Frank, King \& Raine
\cite{FrankKR92}) such that $c_s=h\Omega_K$ where $\Omega_K$ is the
Keplerian angular velocity; hence $c_{s}$ varies radially as
$r^{-1/2}$.  The disk starts from a steady-state velocity profile to
which we add small scale random fluctuations with, unless
otherwise stated, a maximum Mach number of 0.1 (hence the absolute
magnitude of the initial fluctuations also decreases radially).  The
initial steady-state azimuthal velocity is set from the equilibrium
condition, such that the centrifugal force counteracts radial forces
acting on the fluid which include both gravity and gas pressure.
Initially there are no currents within the disk and the magnetic field
is of a purely external origin, being purely vertical with a radial
gradient.  This results in magnetic stresses which are initially
vanishing, but build up with the magnetic potential during the
simulation.  The exact profiles of $B_z$, the initial external field,
and $\Sigma$, the surface density, are varied to examine the criterion
of instability. Linear theory (TP99) lets us expect the disk to
be unstable for initial profiles such that the quantity
$\Omega\Sigma/B_z^2$ increases radially.  This criterion is
investigated in Section \ref{s:gradients}.  The strength of the
external field is also a variable parameter.  The plasma beta, using
the vertically integrated quantities, is defined as
\begin{equation}
\beta=\frac{8\pi P}{B^2} = \frac{8 \pi \Sigma c_s \Omega_K}{B^2},
\end{equation}

\noindent taking into account standard thin disk approximations.  We
vary the field strength using different values of $\beta$ defined at the
inner edge of the disk.  The instability occurs for moderately strong
magnetic fields and is theoretically most active for $\beta\sim 1$.  We
therefore perform tests on $\beta$ in the range $0.1\leq\beta\leq 10$.

\subsection{Non-dimensionalisation}

We choose typical length, time and density scales to
non-dimensionalise the quantities.  We take [l]=$10^8\cm=1000\km$,
[t]=$1000s$ and $[\rho]=10^{-8}\gcmcube$.  These have the added
advantage that velocities are returned as $[v]=1\km\,{\rm s}^{-1}$.
The disk dimensions are then $r=\{1,50\}$, the surface density, we
take to be $\Sigma=10[\rho][\l]=1\kgcmsqr$ and the inner orbital
period is then $T_{r=1}=1.7\times 10^{-4}[t]=0.17\s$.



\section{The numerical methods}

Equations \ref{e:momr} to \ref{e:ind} are solved on a 2d cylindrical
($r,\phi$) grid.  A conservative scheme of the type described by Stone
\& Norman (\cite{StoneN92}) is used to advance the fluid properties in
time.  The scheme uses a staggered mesh such that $\Sigma$ and $B_z$
are cell centered quantities and $u_r$ and $u_\phi$ are edge centered.
This is advantageous in the fact that derivatives, required for a
particular variable, are typically calculated at the desired position
and with increased accuracy.  The grid itself is evenly spaced in
azimuth and logarithmically in radius.  This has two advantages for
this particular model: the region of interest (close to the inner
boundary) has greater resolution and waves propagating to the outer
boundary are damped via inherent numerical dissipation, which
guarantees the outer boundary condition for these simulations
(radiation condition, {\em i.e.} no information coming in from the
outer boundary).  This is particularly important here in reference to
the related Papaloizou-Pringle instability (Papaloizou and Pringle,
1984), which we should recover in the limit of vanishing magnetic
field ($\beta\rightarrow\infty$). In this limit it is well known that
a reflecting outer boundary can result in a much more violent
instability (Narayan {\em et al.}, 1987). We believe that this might
explain the apparently more unstable numerical results of Stehle and
Spruit (\cite{StehleS}).

Averaging of quantities onto the staggered mesh takes into account the
non-uniform nature of the grid.  The resolution is typically taken to be
256x128 cells in $r$ and $\phi$ although lower resolution tests have
been made for comparison.  The high radial resolution is used to capture
the fine details of the instability.  The azimuthal resolution is
arguably excessive since the instability generates low azimuthal
wavenumber spiral waves.  However we found inaccuracies, or even
numerical instabilities, associated with the regularisation of the
logarithmic singularity of the Poisson kernel described by Binney \&
Tremaine (\cite{BinneyT87}) when the azimuthal and radial grid sizes are
very different.  Tests with modified smoothing schemes have been made
with lower azimuthal resolution and indicate that the results presented
here are not dependent upon the scheme, or higher resolution.  Further
details of the Poisson kernel are given in Section \ref{s:numpois}.

\subsection{Evolution of the fluid}

The equations are solved in a conservative form such that angular
momentum, mass and total magnetic flux are conserved to numerical
precision throughout the simulation. The method follows that of Stone
\& Norman (\cite{StoneN92}) such that the terms are split into source
and transport terms: the source terms being centrifugal,
gravitational, pressure and magnetic terms in the momentum equation
and the transport terms being the advection of matter.

The source terms are calculated initially using standard second order
finite difference operators on the staggered mesh (modified for a
logarithmically spaced grid) followed by the transport terms. The
transport terms are calculated using a standard second order Van Leer
(\cite{vanLeer79}) upwind scheme.

We also invoke the FARGO scheme (Masset \cite{Masset00}) to allow for
increased timesteps, typically limited by the underlying large
velocity Keplerian motion. The Courant-Friedrichs-Lewy condition on the
maximal timestep for numerical stability is thus reduced from
$dt\simlt dx/v_{\theta}$ to $dt\simlt dx/v_{g}$, where $dx$ is the grid
size and $v_{g}$ the group velocity of waves which can propagate in the
simulation box. Typically $v_{g}$ will be of the order of the fast
magnetosonic speed, {\em i.e.} comparable with the sound speed if
$\beta\sim 1$. The allowed timestep is thus increased by a factor
$v_{\theta}/v_{g} \sim \epsilon^{-1}$, since $v_{\theta}$ is dominated
by the keplerian velocity.

In fact $v_{g}$ can be estimated more precisely: from the dispersion
relation given by TP99, the maximum group velocity can be approximated
as
\begin{equation}
v_g\sim\frac{v_A^2}{c_s},
\end{equation}
where $v_A=B_z^2\epsilon r/8\pi\Sigma$ is the Alfv\'en velocity.  The
magnetic contribution to the timestep is therefore of the order of
\begin{equation}
dt={\rm min}\left(\Delta x\frac{c_s}{v_A^2}\right),
\end{equation}
taken over the whole of the numerical domain.

\subsection{Boundary conditions}

The effect of boundary conditions on the behavior and development of
the AEI was discussed in detail in TP99.
\subsubsection{Inner boundary}
Concerning the inner boundary, they showed that any reflecting
condition ({\em e.g.} $u_{r}=0$, $u_{\theta}=0$, etc.) results in the
existence of modes with the same instability criterion, but different
frequency. A detailed discussion of the  physics of the inner boundary of
the disk is impossible: indeed, in the example of the accretion disks of
black holes in X-ray binaries, we still do not know what exists between
the disk and the black hole horizon; one could consider an ADAF, or the
magnetic structure threading the black hole, associated with the
Blandford-Zjanek (1977) mechanism. In both cases the detailed physics of
the inner disk boundary is unknown. Hawley and Krolik
(\cite{HawleyK00}) and Hawley(\cite{Hawley00b}) have recently used a
pseudo-newtonian gravitational potential to model the flow of the gas
at the Last Stable Orbit near a black hole. This removes the need for an
artificial boundary condition at the inner disk edge. However the
low-frequency QPO which we consider as a manifestation of the AEI in
X-ray binaries is observed mostly when the disk
stays away from the Last Stable Orbit (Varni\`ere {\em et al.},
\cite{VarniereRTD00}; Rodriguez {\em et al.}, \cite{RodriguezVTD00}).
Here for simplicity we adopt the
commonly used condition $u_{r}=0$.

A disadvantage of this condition is that, since magnetic flux is
advected with the gas, it piles up in the inner few grid-zones.
However, we find that overall, the effective change in the field
strength is acceptable for our current models, where we find that the
plasma $\beta$ changes by 70\% over the course of 200 orbits in the
most extreme case. We feel that, although this is not the most
desirable effect, it is acceptable for the purposes of this paper.
Future models will allow for the magnetic flux to spread in the inner
``hole'', between the disk and the central object.  This will limit
the build-up of the magnetic field (and its gradient) near the inner
disk boundary.

\subsubsection{Outer boundary}

At the outer boundary, reflecting conditions are often used for
simplicity but are non-physical, unless one can argue for the existence
of a sharp outer disk radius.  For instance in the case of an X-ray
binary, a sharp outer boundary may exist, associated with the physics of
tidal effects and the gas flow from the companion.  We have already
mentioned that, as shown by Narayan {\em et al.} (1987), a reflecting
outer boundary may make the instability much more violent.  However in
realistic conditions the ratio of the outer to inner disk radius is very
large, and this effect should not act.  Here the logarithmic radial grid
allows us to extend the simulation grid to very large radius, and there
the grid size also becomes large.  On the other hand for modes of
interest the dispersion relation (Tagger {\em et al.},
\cite{TaggerHSP90}) shows that the radial wavelength becomes small at
large radius.  Numerically, when it becomes comparable with the grid
size, the waves are heavily damped (this is a natural consequence of the
use of low order derivatives and advection).  This ensures that waves
generated in the region of physical interest, at small or moderate
radius, will be damped before they reach the outer boundary, and are
thus not reflected.  This effectively implements the correct boundary
condition that no information should flow from infinity.  A further
advantage of the large radial interval is also that, for a double check,
we can push the outer boundary far enough that waves cannot travel to
the outer boundary within the limited duration of the simulations.

\subsection{Calculation of the magnetic potential}
\label{s:numpois}

As discussed in Section \ref{s:magpot}, the magnetic potential can be
solved analogously to the gravitational potential in a self-gravitating
disk.  For speed of calculation we use the FFT method discussed by
Binney \& Tremaine (\cite{BinneyT87}).  Rather than using the perturbed
density, the perturbed vertical magnetic field is used as the source of
the potential, and the Poisson kernel is modified to account for the
change in coefficients (and sign) between Equations \ref{e:magpois} and
\ref{e:gravpois}.  As in Binney \& Tremaine, we perform a change of
variables to reflect the logarithmic grid and simplify the calculation.
It can thus be shown that after transforming the magnetic field and
potential such that
\begin{eqnarray}
V_M(u,\phi) &=& \exp (u/2)\Phi_M[r(u),\phi] \\
b_z(u,\phi) &=& \exp(3u/2)B_z^D(r(u),\phi)
\end{eqnarray}

\noindent where $V_M$ is the reduced potential, $b_z$ the reduced
perturbed magnetic field and $u$ the logarithm of the radius, $r$,
Equation \ref{e:fullpot} can be rewritten as

\begin{equation}
V_M(u,\phi) = \frac{1}{2\pi}\int_{-\infty}^\infty\int_0^{2\pi}
K(u-u',\phi-\phi')b_z(u',\phi')du'd\phi'
\end{equation}
with the Poisson kernel defined as
\begin{equation}
K(u-u',\phi-\phi') = \frac{1} {\sqrt{2[\cosh(u-u')-\cos(\phi-\phi')]}}.
\end{equation}

\noindent This kernel has a logarithmic singularity where $u-u'=0$ and
$\phi-\phi'=0$.  For this case Binney \& Tremaine (\cite{BinneyT87})
give an approximate smoothing:
\begin{equation}
K(0,0) =
\frac{1}{\pi}\left[\frac{1}{\Delta\phi}
{\rm arcsinh}\left(\frac{\Delta\phi}{\Delta u }\right) +
\frac{1}{\Delta u} {\rm arcsinh}\left(\frac{\Delta u}{\Delta 
\phi}\right)\right]
\label{e:approxK}
\end{equation}

\noindent where $\Delta u$ and $\Delta\phi$ are mesh sizes.  However
this form is only valid for small {\em and comparable} values of $\Delta
u$ and $\Delta\phi$.  Here the physics of the AEI lets us expect (as
confirmed by the results below) only low-$m$ instabilities (where $m$ is
the azimuthal wavenumber), and for speed of computation we often wish to
use much fewer grid cells in $\phi$ than in $u$.  In that case the
approximation (\ref{e:approxK}) gives the unphysical property that
$K(0,0)<K(0,1)$, which in turn generates numerical instabilities.  In
this case we use an alternate method, often used in self-gravitating
disks, modeling the finite thickness of the disk (which does physically
regularise the Poisson kernel).  In order to do this we replace, in the
computation of the kernel, the denominator $|\vec r -\vec r'|$ by
$\sqrt((r-r')^2+h^2$.  This makes the kernel regular at $\vec r=\vec
r'$; it removes the need for a separate evaluation of $K(0,0)$, which
can be written as
\begin{equation}
K(\Delta u,\Delta \phi) =
\frac{1} {\sqrt{2[\cosh(\Delta u)-\cos(\Delta \phi)
+\epsilon^2\exp(\Delta u)]}},
\label{e:modkern}
\end{equation}
where $\Delta u=u-u'$, $\Delta\phi=\phi-\phi'$ and $\epsilon$ is the
aspect ratio of the disk.  For the purposes of the magnetic potential
alone, we have experimented with modifying $\epsilon$ to effectively
alter the smoothing length (however keeping the disk height identical)
and find that it makes only a very slight difference to the extent of
the disk that is affected strongly by the instability.  The overall
development of the instability is unchanged.  This was to be expected
since the smoothing affects only short-range interactions, while the
instability has large wavelengths except at large radii where we wish it
to be damped by the limited resolution of the grid. For the purposes of
this paper, in showing the general characteristics of the instability,
we generally use the Binney \& Tremaine kernel; however we include later
an example of a low azimuthal resolution simulation using the modified
kernel (\ref{e:modkern}).

The power of the FFT method of calculation is that the kernel (and its
Fourier transform) need only be calculated once at the beginning of
the simulation. At the beginning of each time step the Fourier
transform of $b_z$ is calculated.  The inverse Fourier transform of
the convolution of this and the Fourier transform of $K$ is then
performed to generate $V_M$ which in turn gives $\Phi_M$.

\begin{figure*}
\caption{Comparison of the
evolution of instability for different initial values of $\beta=8\pi
p/B^2$, the ratio of thermal to magnetic pressure.  Values of
$\beta=0.1$, $\beta=0.5$, $\beta=1.0$, $\beta=5.0$ and $\beta=10.0$
are shown as dotted, dot-dashed, short-dashed, dot-dot-dot-dashed and
long-dashed lines respectively.  Time is given in units the orbital
time at the inner radius. The plots show the evolution of (top) the
random velocity, and (bottom) the average magnetic field.
\label{f:urmscomp}}
\end{figure*}

The radial and azimuthal magnetic field at the disk surface can then
be calculated from
\begin{equation}
{\bf B} = -{\rm sign}(z)\nabla\Phi_M
\label{e:magpot}
\end{equation}

\noindent yielding $B_r$ and $B_\phi$.  Their jump across the disk
(coming from the sign of $z$ in Equation \ref{e:magpot}) gives the
current in the disk, and thus the Lorentz force in the source terms of
Equations \ref{e:momr} and \ref{e:momphi}.

The initial setup of the field is somewhat unphysical since it assumes
that the field is initially totally of external origin, and we have no
currents in the disk.  The current density in the azimuthal direction is
given by
\begin{equation}
j_\phi=\frac{\partial B_r}{\partial z}-\frac{\partial B_z}{\partial r} =0,
\end{equation}

\noindent therefore, for a magnetic field which is purely vertical at
the disk, and decreases radially one has
\begin{equation}
\frac{\partial B_r}{\partial z}<0.
\end{equation}

\noindent This implies that the field lines must bend inward towards the
central object; in a physically realistic situation the opposite is
usually considered, giving the familiar ``hourglass'' shape commonly
used in disk-jet models.  However, for the purposes of this paper, we
prefer to use this setup which allows us to identify the effect
associated with the field gradient, and leaves out the action of field
curvature and equilibrium currents. Preliminary linear calculations
have shown us that an ``hourglass'' configuration is slightly more
unstable than the one we use here. The code is perfectly able to
describe such initial configurations, with currents both in the central
hole and within the disk. These will be considered in future
publications.



\section{Results}
Our main goal is to characterize the instability, its dependence on the
properties of the disk, and its influence on the global evolution and
accretion rate.  We will thus first vary different parameters affecting
the instability, and its strength.  We will then discuss in selected
cases the detailed properties of the instability, and finally evaluate
its efficiency for accretion.
\subsection{Effect of the field strength}
Let us first consider the influence of the magnetic field pressure in
the disk.  For this we choose equilibrium magnetic field and density
profiles which, according to the linear theory, should give a rather
strong growth rate: the instability criterion (TP99) is that
\begin{equation}
\frac{\partial}{\partial r}\ln \frac{\Omega\Sigma}{B_0^2} > 0.
\label{e:instcriterion}
\end{equation}
We thus take an equilibrium magnetic field varying as
\begin{equation}
B_0(r)=B_0({\rm r_{min}})\left(\frac{\rm r_{min}}{r}\right)^{5/4},
\end{equation}
which is the profile used by Stehle and Spruit (\cite{StehleS}).  This
should be unstable if the surface density profile is flatter than
$\Sigma\sim r^{-1}$.  For this first series of runs we choose a
somewhat unrealistic flat density profile, and vary the parameter
$\beta$. We know that the magneto-rotational instability exists only
for $\beta>1$, whereas the AEI is most unstable for $\beta\sim 1$.

Figure \ref{f:urmscomp} shows the evolution of the rms velocity within
the disk for values of $\beta=0.1, 0.5$, $1.0$, $5.0$ and $10.0$.  As is
shown, the perturbed velocity grows rapidly for $\beta=0.5$ and
$\beta=1.0$, exceeding the initial random noise after approximately 10
orbits.  The growth is approximately exponential during this time.  We
will show in Sections \ref{s:spiral} and \ref{s:spiraltime} that the
instability appears as a spiral mode, in agreement with the theory.  For
the other values of $\beta$ we see very little action in comparison.
The $\beta=0.1$ and $\beta=10.0$ cases are virtually stable with the
$\beta=5.0$ being somewhat more active but failing to achieve the same
amplitude as the most unstable cases.  This run is still interesting to
observe as the growth of the instability is slower and the evolution of
the spiral wave is somewhat cleaner, as presented later.

Also shown in the figure is the rms magnetic field.  This remains
relatively constant over time for all cases, with only moderate
increases in $\beta$ occuring for the unstable cases ({\em i.e.}
$\beta=0.5$, 1.0 and 5.0) as shown in
Figure~\ref{f:betacomp}. The maximum change in $\beta$ occurs for the
$\beta=1.0$ case. We see it decreasing to approximately $\beta=0.6$
after the initial development of the instability as the field and
matter are accreted to the inner boundary. After this time we see some
oscillatory features and finally a very gradual decrease to a value of
$\beta\sim0.3$ at around 200 orbits. Hence the plasma beta decreases
to approximately 30\% its initial value for the most active case.

In the subsequent sections we examine more closely the most active
values of plasma beta (namely $\beta=0.5$, $\beta=1.0$ and
$\beta=5.0$).  As seen from the growth and development of the
perturbed velocity, the cases $\beta=0.5$ and $\beta=1.0$ are similar
and this holds true for many other characteristics in the structural
evolution. Similar features hold also for the $\beta=5.0$ case with
development being somewhat slower.

\begin{figure}
\begin{center}
\caption{Evolution of $\beta$ at the inner edge of the disk for the four
runs starting from $\beta=0.5$ (dotted), $1.0$ (solid),
$5.0$ (dashed) and $10.0$ (dot-dashed). Time is given in
terms of the orbital period of the inner radius.
\label{f:betacomp}}
\end{center}
\end{figure}

\begin{figure}
\caption{Comparison of runs with
different azimuthal resolutions and approximations of the Poisson
kernel, for $\beta=1.0$.  The radial resolution is fixed at 256 grid
zones.  The dashed curve shows the results of the run with 128 azimuthal
grid zones and the standard kernel; dotted is for 32 zones and the
smoothed kernel.
\label{f:rescomp}}
\end{figure}

\subsection{Effect of the spatial resolution}

As discussed in Section \ref{s:numpois}, the Poisson kernel given by
Binney \& Tremaine (\cite{BinneyT87}), and used in the computation of
the magnetic potential, is only valid for high azimuthal resolution
(comparable with the radial one).  For this reason the simulations
presented in the majority of this paper are for $256 \times 128$ grid
zones in $u$ and $\phi$ respectively.  The high radial resolution is
needed to obtain a good description of the AEI, which depends critically
on a resonance localised at the corotation of the modes (where their
angular phase velocity equals the rotation velocity of the gas).  On the
other hand the instability is predicted to occur only for very low
azimuthal wavenumber $m$, so that a high angular resolution would be
necessary only to describe in details the dissipation of wave
energy at small scales, as the non-linear evolution leads to wave
steepening an possibly to the formation of a spiral shock.  This goes
far beyond the purpose of the present paper and would require much more
elaborate simulations.

Thus simulations with low angular resolution, and increased radial
resolution, would be enough for our present purposes.  As discussed in
Section \ref{s:numpois}, this leads to a numerical instability if we use
the regularisation, given by Binney \& Tremaine (\cite{BinneyT87}), of
the logarithmic singularity of the Poisson kernel; we have made low
azimuthal resolution runs with a different regularisation, corresponding
to a smoothing by the finite vertical thickness of the disk.  In this
section we compare two simulations: one using the standard Binney \&
Tremaine kernel at $256 \times 128$ and another with a smoothed kernel
(assuming a disk aspect ratio $\epsilon=0.1$) with a resolution of
$256\times32$.  All other parameters are identical for the two runs,
with $\beta=1.0$.

Figure \ref{f:rescomp} shows the different evolutions of the two
cases.  As seen, the initial evolution of the instability is virtually
identical.  A similar magnitude of $u_{rms}$ is achieved after
approximated 20 orbits. Only the final, saturated state of the
instability (which depends at least in part on the dissipation of wave
energy at small scales) is slightly affected, both for the perturbed
velocity and magnetic fields.  This confirms that the smoothed kernel
does give the correct physics we expect.


\subsection{Effect of the radial density gradient}
\label{s:gradients}

We now retain the same radial magnetic field profile, but change the
density profile to check its effect on the growth of the instability.
We have chosen not to try and reproduce the profiles used in the
linear analysis of TP99: these profiles, imposed by the numerical method
used in TP99 to solve the linear system, have a sharp and localised
gradient at the corotation radius, in order to limit the range of
density and/or magnetic field in the disk.  They would thus be smeared
rapidly by the non-linear evolution of the instability, precluding the
more direct comparison of theory with numerical simulation one could
expect.  Work in progress with a different scheme for the solution of
the linear problem should relax this condition, and allow a direct
comparison of the theory with the non-linear simulations.
\begin{figure}
\caption{Comparison of runs with different density gradients, for
$\beta=1.0$ at the inner edge of the disk.  Lines plotted are dashed:
$\Sigma={\rm const}$ for a $256\times128$ grid, dotted: $\Sigma\propto
r^{-3/2}$ for $256\times128$, dot-dashed: $\Sigma\propto r^{-3/2}$ for
$512\times512$ and dot-dot-dot-dashed: $\Sigma\propto r^-2$ for
$512\times512$.
\label{f:gradcomp}}
\end{figure}

\begin{figure*}
\newlength\ltest
\ltest=\hsize
\divide\ltest by 3
\caption{Evolution of the spiral structure at regular intervals in time
for the $\beta=5$ case.  Starting from an initial random state the
instability structure evolves from an $m=3$ (three-armed) spiral to an
$m=2$ and an $m=1$.  The plots show the radial velocity.  Times shown
are in rotation periods at the inner disk edge.
\label{f:b=5.sprl}}
\end{figure*}

To study the stability criterion we chose two extra cases: one
with a surface density varying as $\Sigma\sim r^{-3/2}$ and a steeper
one with $\Sigma\sim r^{-2}$.  These should be increasingly stable.
The first case is identical to that chosen by Stehle \& Spruit.
According to the criterion (\ref{e:instcriterion}), only the flat
density profile should be subject to the AEI.

Figure \ref{f:gradcomp} displays the growth of the instability for these
three different gradients of surface density, along with a run of
different resolution.  The initial random velocities vary slightly since
the deviation from Keplerian rotation (thus the effect of the pressure
gradient on the equilibrium rotation curve) is included in the rms
values.

This shows that, as expected, a radial surface density gradient
stabilizes the modes.  For $\Sigma\sim r^{-2}$ the instability has
totally disappeared.  For $\Sigma\sim r^{-3/2}$ it is still present,
though weaker than with a flat density profile.  Thus numerically the
instability threshold seems to be steeper (in terms of the density
profile) than given by the criterion (\ref{e:instcriterion}).  This
can be attributed to the fact that, as discussed by TP99, there are
actually two different contributions to the growth rate: the
corotation resonance gives the criterion (\ref{e:instcriterion}).
However a second mechanism (actually the first discovered by Tagger
{\em et al.}, \cite{TaggerHSP90}) corresponds, as for galactic
spirals, to the emission of an outgoing spiral wave, emitted beyond
the corotation radius.  This outgoing spiral is very visible in
the plots, figures 5-7. This mechanism does not depend on the radial
profiles (we refer to TP99 for a more complete discussion).  Although
we expect it to be weaker than the corotation resonance, it shifts the
instability threshold leaving the $\Sigma\sim r^{-3/2}$ case
weakly unstable.  The presence of the outgoing spiral will be
discussed later in more details.  In order to verify the instability
of the $\Sigma\sim r^{-3/2}$ profile, we have performed a very high
($512\times 512$) resolution run, giving results similar to the lower
resolution one (though the growth rate is slightly lower, as shown in
Table \ref{t:growth}).

\subsection{Growth rates}

The plots of rms velocity (Figures \ref{f:urmscomp}, \ref{f:rescomp} and
\ref{f:gradcomp}), show that in general the instability grows
exponentially until it reaches a saturated state.  We give in Table
\ref{t:growth} the growth rates in terms of $\Omega_0$ - the
angular velocity at the inner radius - of all the runs presented here
covering the different resolutions, $\beta$ and radial density
gradients.  The growth rate indeed appears to be strongest for the flat
density gradient with $\beta=1.0$, and to almost totally disappear when
$\beta$ is one order of magnitude higher or lower. Similarly, as
the density gradient gets steeper, the growth rate decreases and the
case with $\Sigma\sim r^{-2}$ is totally stable.

\begin{table}[!ht]
\begin{center}
\begin{tabular}{cccc}
\hline
$\beta$ & Resolution & $a$ ($\Sigma\propto r^{-a}$) & Growthrate/$\Omega_0$ \\
\hline
\hline
0.1 & $256\times128$ & 0 & 0.0001 \\
0.5 & $256\times128$ & 0 & 0.0346 \\
1.0 & $256\times128$ & 0 & 0.0369 \\
5.0 & $256\times128$ & 0 & 0.0058 \\
10.0 & $256\times128$ & 0 & 0.0002 \\
\hline
1.0 & $256\times32$ & 0 & 0.0371 \\
\hline
1.0 & $256\times128$ & 3/2 & 0.0173 \\
1.0 & $512\times512$ & 3/2 & 0.0143 \\
1.0 & $512\times512$ & 2 & 0.0000 \\
\hline
\end{tabular}
\end{center}
\caption{Comparison of growth rates for all runs performed.
\label{t:growth}}
\end{table}

These values are comparable with the theoretically predicted ones, as
given in TP99.  As explained in the previous section, a detailed
comparison is difficult, since TP99 use quite unrealistic radial
profiles, in order to limit the difficulties associated with their
solution method.

\subsection{Evolution of the spiral wave}
\label{s:spiral}

The evolution of the instability into a well developed spiral wave is
best seen for the $\beta=5.0$ case. Figure \ref{f:b=5.sprl} shows the
evolution from the initial random velocity field at intervals of 20
orbits. The plots are of the radial velocity in the inner region of
the disk.

By $t=60$ orbits (at the inner edge of the disk), an $m=3$ (3-armed)
spiral is clearly evident.  This then develops over time to lower
wavenumbers, eventually becoming a tightly wound $m=1$ spiral after
about 200 orbits.  The disk goes through some intermediate stages for
which a mixture of wavenumbers co-exist.  This is discussed in more
detail in the following section.

The development of the spiral from higher to lower wavenumbers is again
in agreement with the theory: the most unstable azimuthal wavenumber
decreases from $m=$ a few to $m=1$ as $\beta$ decreases. This decrease
of $\beta$, shown in Figure \ref{f:betacomp}, occurs as magnetic flux is
advected inward with the gas to the inner region of the disk.

All the runs show similar behaviour to the $\beta=5.0$ case. The
spiral always develops from high to small wavenumbers, eventually
becoming $m=1$ in all cases.

\subsection{Time analysis of the development of the instability}
\label{s:spiraltime}

\begin{figure}
\ltest=\hsize
\caption{Contour plot of radial velocity as a function of radius and
time, at $\phi=0$, in the inner region of the disk for the
$\beta=1.0$ case. Oblique features show a wave travelling outward,
while horizontal ones indicate a standing pattern.
\label{f:b=1.urt}}
\end{figure}
\begin{figure}
\caption{Outgoing waves propagating outward.  Radial velocity is plotted
radially at $\phi=0$ , in the $\beta=0.5$ run.  The times plotted are
$t=100.0$ (solid), $t=100.2$ (dashed), $t=100.4$ (dot-dashed), $t=100.6$
(dot-dot-dot-dashed), $t=100.8$ (dotted).
\label{f:b=0.5.ogwaves}}
\end{figure}
\begin{figure}
\caption{Vorticity in the inner disk region in the $\beta=0.5$ run.
\label{f:b=0.5.vort}}
\end{figure}

\begin{figure}
\caption{Velocity field in the inner disk region, in the case shown in Figure
\ref{f:b=0.5.vort}.  Large azimuthal velocities close to the inner edge
have been removed from the plot for clarity.
\label{f:b=0.5.2dflow}}
\end{figure}

\begin{figure}
\caption{The same flow pattern as in Figure \ref{f:b=0.5.2dflow}, in a
Cartesian projection.  The vortex is clearly visible around $r\sim 1.5$.
\label{f:b=0.5.vortic2}}
\end{figure}

We further this analysis of the evolution of the instability by
performing a spectral analysis of the development of its
structure with time.

Figure \ref{f:b=1.urt} shows a contour plot of the radial velocity,
measured on the axis $\phi=0$, as a function of radius and time, for the
$\beta=1$ run, with lighter regions indicating higher velocities.  This
provides some essential details on the nature of the instability.  It
becomes visible after $t\sim20$ orbits.  Higher velocities are seen to
occur close to the inner radius of the disk.  In the innermost region
($r\simlt 2$), the contours shown are horizontal, whereas farther out
they are oblique with a positive slope: the latter represent a wave
travelling outward, while the former indicate a standing pattern.  This
is the structure expected for the AEI which (like self-gravity driven
galactic spirals) distinguishes two regions in the disk, separated by
the corotation radius (where the angular phase velocity of the wave
equals the rotation frequency of the gas): inward from corotation, the
instability establishes a standing pattern formed of an outgoing and an
ingoing spiral.  Beyond corotation, it emits a trailing spiral wave
travelling outward.

This all occurs very close to the centre of the disk.  Virtually
no effects are observed for $r$ much greater than $20$.  This is also a
good indication that the outer boundary will have little effect on the
results.  Virtually no waves travel to the outer boundary and hence
reflections and subsequent interactions can be neglected. Any
significant signal returning from the outer boundary would appear in
this plot as an oblique pattern with a negative slope.

Figure \ref{f:b=0.5.ogwaves} shows in more details the outgoing
character of the wave in the outer region.  Figure \ref{f:b=0.5.vort}
shows the vorticity in the flow.  A sharp feature is seen at $r\simeq
1.5$.  This again is expected from the AEI, which grows by coupling the
spiral pattern to a Rossby vortex it generates at its corotation radius.
The sharp vorticity feature is a manifestation of this vortex.  This
will play a crucial role in future studies: the instability mechanism
lies in the fact that the spiral inward from corotation extracts energy
and angular momentum from the flow (thus causing accretion), and
transfers them either to the Rossby vortex or to the outgoing spiral
beyond corotation.  The vortex is usually more efficient (as discussed
in Section \ref{s:gradients}).  It can be considered as a twisting of the
footpoints of magnetic field lines in the disk.  Thus, if we took into
account a small but finite density above the disk, this twisting would
propagate as Alfv\'en waves, carrying vertically toward the corona (where
they could power a jet or an outflow) the energy and angular momentum
extracted from the disk. This is the process which inspired the name of
the Accretion-Ejection Instability.

In order to show the vortex more clearly, we have plotted in Figure
\ref{f:b=0.5.2dflow} the flow pattern in the inner region of the disk.
A turning point in the flow is seen close to the inner boundary in the
top-right quadrant of the plot.  This vortex is more clearly shown in
the flow pattern of Figure \ref{f:b=0.5.vortic2}.  For clarity, both
figures show the deviations from the underlying Keplerian flow.  A flow
pattern around a singular point is seen in both plots.  This point is at
the corotation radius.

Figure \ref{f:b=5.specs} shows details of the spatial structure of the
waves during four intervals of time, for the $\beta=5.0$ run, which
shows most clearly the spectral evolution.  For each time interval it
shows two plots: the left one is a contour plot of the radial velocity,
analogous to the one shown in Figure \ref{f:b=1.urt}.  The right plot
shows contours of the Fourier transform of the left one, giving as a
function of radius, the frequencies at which perturbations evolve during
the given time interval.  Frequencies of $\omega/\Omega_{0}\simeq 2.2,\
1.5\ {\rm and\ } .7$ appear succesively, corresponding to the $m=3, \ 2
\ {\rm and\ } 1$ spirals seen in Figure \ref{f:b=5.sprl}.  These
frequencies correspond to corotation radii which increase ($\omega/m$
decreases) as $m$ decreases, as expected from the theory.

We note that results recently released by Hawley \& Krolik
(\cite{HawleyK00}) and Hawley (\cite{Hawley00b}) show similar
behaviour close to the inner edge of an accretion disk.  Their
numerical setup is much more complex than ours, and describes a 3D MHD
model of the inner region of an accretion disk around a black hole
with a pseudo-Newtonian potential.  Although their goal is to study
the magneto-rotational instability in a realistic 3D geometry, they
note that after some time a significant part of the radial transport
is done by waves.  They show, in the inner region of the disk, plots
very similar to our Figure \ref{f:b=1.urt}.  A detailed comparison is
difficult, since this appears after an important evolution of the disk
and magnetic field geometry.  Furthermore, the relatively poor radial
resolution imposed by the full 3D simulation would probably rule out a
proper description of the AEI. However the similarity between their
results and our Figure \ref{f:b=1.urt} (including the horizontal and
oblique patterns discussed previously) is striking.  We also note
that, when this appears, the magnetic field has strongly evolved from
their initial setup, as some vertical magnetic flux has been advected
inward with the gas.  In the inner region of the disk, they thus have
an essentially vertical field of rather high amplitude (although the
value of $\beta$ is not given).  We thus consider that the wave
patterns they obtain could very well be a manifestation of the AEI,
rather than waves generated non-linearly by the magneto-rotational
turbulence.
\begin{figure*}
\ltest=\hsize \divide\ltest by 2
\caption{Contour plot of the radial velocity against time close in the
inner disk region, in the $\beta=5.0$ run, with associated spectral
analysis.  Time ranges shown are for 50 to 75 orbits, 100 to 125 orbits,
150 to 175 orbits and 200 to 225 orbits.  Spectra show first a feature
at $\omega/\Omega_{0}\simeq 2.2$, then one at $\simeq 1.5$, and finally
at $\simeq .7$.  They correspond to the $m=3, \ 2\ {\rm and}\ 1$ spirals
succesively formed in the disk.
\label{f:b=5.specs}}
\end{figure*}
\subsection{Accretion rates}

 From the simulations we are able to estimate the efficiency of the
instability in accreting matter towards the central object. At a given
radius the accretion rate is calculated as:
\begin{equation}
\dot{M}(r) = - \int_0^{2\pi}r u_r(r,\theta)\Sigma(r,\phi) d\phi.
\end{equation}

\noindent However, this value is highly variable.  Plotting against
$r$ at one point in time shows very little of the overall effect of
the instability.  We therefore show values that are averaged over
space or time, in order to obtain a better understanding of the
overall accretion within the disk.  All figures shown are for the
$\beta=1.0$ run.

Figure \ref{f:b=1.acc} shows the accretion rate against time, averaged
between $r=1$ and $r=10$.  This region is chosen since the instability
has little effect at larger radii.  This shows that even though the
accretion rate is highly intermittent, it is predominantly positive.
Three bursts of accretion are seen before the first 100 orbits, at
$t\sim 30$, $t\sim50$ and $t\sim75$ orbits.  These burst are also seen
in the plots of rms velocity in Figure \ref{f:urmscomp}.

However, the physics of the instability lets us expect a more complex
behavior: spiral density waves carry a {\em negative} flux of energy and
angular momentum inside their corotation radius, and a {\em positive}
one beyond corotation: this means that a wave inside corotation grows by
extracting energy and angular momentum from the disk, and transfering
them either to the Rossby vortex, at the corotation radius, or to the
outgoing spiral, beyond it.  Accordingly, we expect to see (at least
during the initial phase when the instability grows following its linear
behavior) accretion in the inner part of the disk, and outward motion
(negative $\dot M$) beyond the corotation radius, which is typically of
the order of $1.5 - 2$.  This is exactly what we see in Figure
\ref{f:inneraccretion}a, which shows the accretion flux in the inner
region of the disk, averaged between times 10 and 25 and for the same
run with $\beta=1$.  However the result is still quite noisy, especially
at large radius: this turns out to come from our initialization, where
we have introduced strong ($\sim .1$) random fluctuations of the radial
velocity, from which the instability grows.  By the time these
fluctuations have dissipated or traveled away, they have caused
significant radial flows.  In order to check this we have performed a
new simulation with the initial rms noise reduced to $.001$: Figure
\ref{f:inneraccretion}b shows that the noise has strongly decreased at
large radius, while the inner region is qualitatively similar (note that
absolute values of $\dot M$ cannot be compared, since the instability
now takes longer to grow from the initial conditions to a given
amplitude).  Finally we run a new case where the initial random
perturbation now depends on $r$ as a Gaussian centered around $r=5$:
Figure \ref{f:inneraccretion}c shows the resultant accretion flow, which
has the same structure as in Figure \ref{f:inneraccretion}b, but where
the fluctuations at large radius have totally disappeared.  Figure
\ref{f:inneraccretion}d is a zoom on the inner disk region, showing that
as expected $\dot M$ is positive inside a radius $r\simeq 1.7$, and
negative beyond, going to zero at large radius.

At larger times, however, accretion becomes positive beyond the
corotation radius: this is shown by Figure \ref{f:b=1.acc.gauss}, where
$\dot M$ is averaged over the whole simulation.  It is interesting to
see that, in the region between corotation and $r=10$, most of this
positive accretion occurs by bursts.  These (already noted in Figures
\ref{f:urmscomp} and \ref{f:b=1.acc}) are shown on Figure
\ref{f:b=1.Mdotrt}, where $\dot M$ is shown as a function of $r$ and
$t$.  The bursts are seen as strong streaks propagating outward from the
corotation region.  We conclude that this is due to non-linear effects
({\em e.g.} shocks, or global re-arrangements of the current profile)
not included in the linear theory, but responsible for the saturation of
the instability after a finite time.  We defer their full
characterization to future work where more realistic magnetic field and
current profiles will be used.
\begin{figure}
\caption{Accretion rate,
averaged between $r=1$ and $r=10$, against time for the $\beta=1.0$ run.
\label{f:b=1.acc}}
\end{figure}

\begin{figure*}
\ltest=\hsize \divide\ltest by 2
\caption{The accretion rate $\dot M$ as a function of radius, for the
$\beta=1$ run, in the linear growth phase of the instability.  Results
are shown from runs with different initial conditions for the random
velocity fluctuation in the disk.  a (top left): strong and constant
(constant meaning maximum Mach number), b (top right) weak and
constant, c (lower left) weak, peaked around $r=5$, and d (lower
right) magnification of the inner region of c.  As expected for spiral
waves in the disk, accretion is positive inside the corotation radius,
and negative outside.
\label{f:inneraccretion}}
\end{figure*}

It is customary to measure the efficiency of accretion in terms of a
turbulent viscosity, parametrised with the parameter $\alpha_{ss}$ of
Shakura \& Sunyaev (\cite{ShakuraS73}).  This allows a comparison of
the instability mechanism with standard $\alpha$-disk model.  We thus
convert the accretion rate to an equivalent value of $\alpha_{ss}$,
although we emphasize that the physics of the AEI (which forms
large-scale, quasi-stationary patterns where the accretion energy is
carred away by waves) is very different from the assumptions
underlying the Shakura-Sunyaev model, {\em i.e.} small-scale
turbulence leading to {\em local deposition} of the accretion
energy. Here, as explained in previous sections and in more
details in TP99, the instability mechanism relies essentially on the
excitation of the Rossby vortex by the spiral wave. This occurs
through the long-range action of the magnetic potential
$\Phi_{M}$. Thus the accretion energy is not deposited locally, but
transported at large distances by the magnetic stresses figuring in
equations (1-2). The present estimate can thus be taken only as a
convenient measure of the efficiency of the instability to cause
turbulent transport.  Furthermore, a reliable assesment of the
accretion rate due to the AEI would require taking into account the
vertical emission of Alfv\'en waves ({\em i.e.} include the effect of
a finite density corona above the disk), and effects associated with
the finite thickness of the disk.

 From standard thin disk approximations of steady accretion disks, one
has
\begin{eqnarray}
\nu\Sigma &=&
\frac{\dot{M}}{3\pi}\left[1-\left(\frac{r_*}{r}\right)^{1/2}\right],
\nonumber \\
&\sim& \frac{\dot{M}}{3\pi},
\label{e:fkr92}
\end{eqnarray}

\noindent (Frank, King \& Raine \cite{FrankKR92}) where $\nu$ is the
viscosity, $\Sigma$ the surface density and $r_*$ is the radius of the
central object. The standard prescription for the viscosity is given
by
\begin{equation}
\nu=\alpha_{ss} c_s h.
\label{e:alpha}
\end{equation}

\noindent Combining Equations \ref{e:fkr92} and \ref{e:alpha} along
with another thin disk assumption, that $c_s=h \Omega$, yields
\begin{equation}
\alpha_{ss}(r,t) = \frac{\dot{M}(r,t)}{3\pi \epsilon u_\phi(r,t)r\Sigma(r,t)},
\end{equation}

\noindent where $\epsilon$ is the disk aspect ratio.  The values of
$u_\phi(r,t)$ and $\Sigma(r,t)$ represent azimuthally averaged
quantities resulting in an $\alpha_{ss}$ that depends on radius and
time.

Figure \ref{f:b=1.alpha} shows the effective $\alpha_{ss}$ against time
when averaged between $r=1$ and $r=10$, for the $\beta=1$ run with
Gaussian initial perturbations.  It is negative while the wave
is in its linear growth phase, but becomes positive in the non-linear
stage. It is again strongly intermittent, showing the same bursts as in
previous figures.

It is useful to compare values of $\alpha_{ss}$ with values derived
from numerical models of turbulent accretion disks, for which the use
of the $\alpha$ parameter is more applicable. The use of shearing-box
models ({\em e.g.} Brandenburg {\em et al.} \cite{BNST95}, Brandenburg
{\em et al.}  \cite{BNST96}, Stone {\em et al.} \cite{StoneHGB96},
Hawley {\em et al.}  \cite{HawleyGB96}) over recent years have given
insights to the turbulent transport of energy via calculations of
Reynolds and Maxwell stresses as well as other methods (such as from
the overall increase in thermal energy). Various values of $\alpha$
calculated from different quantities by Brandenburg {\em et al.}
(\cite{BNST96}) indicate that their $\alpha_{ss}$ is typically of the
order of 0.001 to 0.01 with maxima generally around 0.01 to
0.02. These are comparable to those of Stone {\em et al.}
(\cite{StoneHGB96}) and and Hawley {\em et al.}  (\cite{HawleyGB96}).

With the AEI simulation in 2d for $\beta=1.0$ we see similar values of
the viscosity parameter, $\alpha_{ss}$. The averages over time and
space yield values typically between 0.001 and 0.006. Maximum values
at a given position and time are around 0.06. These are still somewhat
smaller than the recent values given by Hawley \& Krolik
(\cite{HawleyK00}) for their 3d model of the inner accretion disk and
also that of Armitage (\cite{Armitage98}).

\begin{figure}
\caption{The accretion rate $\dot M$ as a function of radius, for the
$\beta=1$ run, averaged over the whole simulation. Accretion has become
positive up to $r\simeq 10$.
\label{f:b=1.acc.gauss}}
\end{figure}
\begin{figure}
\caption{Evolution of radial structure of accretion rate for $\beta=1$ with
initial perturbations given a Gaussian profile.
\label{f:b=1.Mdotrt}}
\end{figure}

\begin{figure}
\caption{Estimation of an equivalent Shakura-Sunyaev $\alpha$-parameter from
the accretion rate, averaged between $r=1$ and $10$, in the $\beta=1$
run.
\label{f:b=1.alpha}}
\end{figure}

The strength of the instability, depending on $\beta$ and other
parameters, affects the value of $\alpha_{ss}$. Table
\ref{t:accr} shows the maximum values of $\alpha_{ss}$ from each of
the runs.

\begin{table}[!ht]
\begin{center}
\begin{tabular}{cc}
\hline
$\beta$ & $\alpha_{ss}$ \\
\hline
\hline
0.1 & 0.011 \\
0.5 & 0.122 \\
1.0 & 0.056 \\
5.0 & 0.030 \\
10.0 & 0.013 \\
\hline
\end{tabular}
\end{center}
\caption{Comparison of equivalent values of $\alpha_{ss}$ derived from
the accretion rate for each case of $\beta$.
\label{t:accr}}
\end{table}

Again we see that the instability is strongest for moderate values of
$\beta$, {\rm i.e.} around $\beta\sim 0.5$. Indeed, we see that values
of $\alpha$ are strongly peaked for $\alpha=0.5$ and are more
comparable to values desirable for models of cataclysmic variables
(Cannizzo \cite{Cannizzo93}). Recent global simulations of the MRI
(Armitage \cite{Armitage98}, Hawley \cite{Hawley00}) indicate that
values of $\alpha$ derived from global numerical simulations are
generally higher than local shearing-box simulations ({\em e.g.}
Brandenburg {\em et al.} \cite{BNST96}, Stone {\em et al.}
\cite{StoneHGB96}, Miller \& Stone \cite{MillerS00}) and hence the
possibility exists of cataclysmic variables occuring as a result of a
combination of both the AEI and the MRI. We must also emphasize that
the physics discussed here concerns only the innermost part of the
disk. Presumably, if accretion proceeds faster at larger radii, the
resulting inflow of matter and magnetic flux will strongly affect the
evolution of the AEI and its consequences on inward transport.



\section{Conclusions}

In this paper we have presented a two-dimensional model of the inner
region of an accretion disk threaded by an initially vertical magnetic
field, surrounded by vacuum.  Starting from an equilibrium situation
with additional random small scale fluctuations, an instability is seen
to occur on a timescale of a few tens of orbits.  The instability
appears as a low azimuthal wavenumber spiral wave, amplified mainly by
(and feeding) a Rossby vortex.  These results are comparable (with a
slightly different physical setup and a more adapted numerical one) with
the ones obtained recently by Stehle and Spruit (\cite{StehleS}).

The instability is seen to occur most strongly for a plasma beta of
$\beta\sim 1$ ({\em i.e.} equipartition between the gas thermal
pressure and magnetic pressure in the disk), with its amplitude and
growth rate rapidly decreasing for both stronger and weaker fields.
Also, the growth rates (and the existence of the instability) depend on
the radial gradients of equilibrium quantities.

As the spiral wave develops and affects the disk structure (in
particular decreasing its $\beta$ value) it is seen to progress from
larger to smaller azimuthal wavenumbers until it reaches $m=1$, {\em
i.e.} a one-armed spiral.

These properties (timescales, development, characteristics, stability
criterion) allow us to identify the instability as the
Accretion-Ejection Instability, and globally to confirm qualitatively
and quantitatively the behaviours predicted theoretically by TP99.

The instability is seen to generate accretion in the inner region of the
disk.  The amplitude of the accretion -- calculated through an equivalent
value of the standard $\alpha$ parameter of viscous transport (Frank,
King \& Raine \cite{FrankKR92}) -- is comparable, when compared to other
numerical simulations of accretion disks ({\em e.g.} Brandenburg {\em et al.}
\cite{BNST95}, Stone {\em et al.} \cite{StoneHGB96}), with the one generated at
lower magnetic pressure by the magneto-rotational instability (MRI) of
Balbus \& Hawley (\cite{BalbusH91}) .

Furthermore, we have noted that waves observed in more recent
publications of global, 3D simulations of the inner region of
accretion disks (Hawley \& Krolik \cite{HawleyK00}, Hawley
\cite{Hawley00b}) show properties (development of spiral waves with a
characteristic propagation pattern) which are very similar to the ones
described here.  These waves, which seem to cause a significant part
of the accretion, develop once the global magnetic structure has
evolved in the inner region of the disk to become quite similar to the
one used here. It is thus quite possible that the AEI is also present
in these simulations, once the Magneto-Rotational Instability has
caused the disk to evolve to a favorable magnetic configuration (in
terms of geometry and strength) in its inner region.

Future work needs to extend the current simulations to 3D, in two steps:
the first one will be to consider a disk which for simplicity is still
infinitely thin, but embedded in a corona of small but finite density:
this will allow us to describe the vertical ejection, as Alfv\'en waves,
of the energy and angular momentum extracted from the disk and
presently stored in a Rossby vortex. It will thus be possible to
consider how they can ultimately be deposited in the corona, where they
could energize a jet or an outflow.

A second step will be to turn to fully 3D simulations. This leads to
numerical constraints which are quite different from the ones
encountered in the simulation of magneto-rotational turbulence, since
the physics of the AEI is quite different ({\em e.g.} it needs very few
grid points within the disk in the vertical direction, but a large
number in the radial one). This will allow us to address the
acceleration of the gas, across a slow magnetosonic point, to finally
form a jet.

\begin{acknowledgements}

The authors gratefully acknowledge the help of F. Masset in
providing valuable support in the development of many of the numerical
methods described in this paper. They also acknowledge many discussions
with R. Stehle and H. Spruit on their comparable simulations.

\end{acknowledgements}


\end{document}